\documentclass[12pt]{iopart}
\usepackage{graphicx}
\usepackage{setstack,cite}
\begin{document}
\date{}

\title[Complex Nonlinear Oscillators]
{A nonlocal connection between certain linear and nonlinear ordinary
differential equations - Part II: Complex nonlinear oscillators}

\author{R. Mohanasubha, Jane H. Sheeba, V. K. Chandrasekar, M.~Senthilvelan and M. Lakshmanan}%
\address{Centre for Nonlinear Dynamics, Department of Physics,
Bharathidasan University, Tiruchirappalli - 620 024, India }

\date{\today}

\begin{abstract}
In this paper, we present a method to identify integrable complex nonlinear oscillator systems and construct their solutions. For this purpose, we introduce two types of nonlocal transformations which relate specific classes of nonlinear complex ordinary differential equations (ODEs) with complex linear ODEs, thereby proving the integrability of the former. We also show how to construct the solutions using the two types of nonlocal transformations with several physically interesting cases as examples. 
\end{abstract}


\section{Overview}
\subsection{Motivation}
In a previous paper \cite{Chand:07:nld} we have developed a new way of identifying integrable nonlinear ordinary differential equations (ODEs) by relating linear and nonlinear oscillator equations of any order through suitable nonlocal transformations. We have also devised a method to derive explicit general solution of the nonlinear ODEs. Interestingly the nonlocal oscillators identified through this procedure posses certain remarkable properties \cite{Chand1}. In this paper we extend the underlying features of this method to the case of complex ODEs and identify a class of integrable complex nonlinear oscillators including
hierarchy of Stuart-Landau equations, complex modified Emden equation, complex Duffing-van der Pol oscillator equation and so on.

The necessity of this analysis comes from the fact that complex nonlinear ODEs are being used to describe autoresonance/parametric resonance phenomenon in a number of problems in different branches of physics and are being widely used in the contemporary nonlinear dynamics literature \cite{Kuramoto:84,Pikovsky:01}. Some of the interesting systems include, the Poincar\'e equation $\frac{dz}{dt}+\alpha z+\beta |z|z=0$ and 
Stuart-Landau oscillator equation $\frac{dz}{dt}+\alpha z+\beta |z|^2z=0$ which play crucial roles in explaining complex network properties of systems in physics,
chemistry, biology and sociology, where coupled models of these systems are of basic interest. Examples include Josephson junction arrays, lasers, coupled neural networks, pacemaker cells in the heart, ensembles of electrochemical oscillators and so on. Investigating the integrability properties and exploring the underlying solutions of these complex nonlinear ODEs help one to understand the basic dynamics of these nonlinear systems.

Moreover, in recent times PT-symmetric complex Hamiltonians in the classical cases, satisfying complex ordinary differential equations for the dynamical variables, are being intensively studied \cite{ben1} and their quantum versions investigated \cite{ben2}. In such a situation the study of complex nonlinear ODEs becomes very crucial. 
The foremost question now becomes how to identify integrable complex nonlinear ODEs of arbitrary order and explore their general solutions in a systematic way. The question becomes easier when one has a specific equation. In this case one may rewrite the given complex nonlinear ODE as a set of real ODEs by appropriately splitting the complex variable into real variables and then start to investigate the integrability of these real ODEs through any one of the analytical or geometrical methods, namely Painlev\'e test \cite{mlsrb}, Lie symmetry analysis \cite{blum}, Prelle-Singer procedure \cite{Chand2}, Darboux method \cite{libre} and so on. In this paper we extend the idea given in our previous paper \cite{Chand:07:nld} for the case of real ODEs to the case of complex ODEs and identify the integrable equations in a direct manner using nonlocal transformations.
\subsection{Earlier work}
Let us recall our earlier work \cite{Chand:07:nld} briefly here.
Consider a linear harmonic oscillator equation, $\ddot{U}+\lambda U=0$, where over dot denotes differentiation with respect to $t$ and $\lambda$ is a real arbitrary parameter, and introduce a nonlocal transformation $U=x(t)e^{\int_0^t{f(x(t'))}dt'}$, where $f(x(t))$ is an arbitrary function of $x$. Substituting  this nonlocal transformation (NLT) into the linear equation one gets a nonlinear ODE of the form
\begin{eqnarray}            
\ddot{x}+(2f+xf')\dot{x}+(f^2+\lambda)x=0,
\label {int03}
\end{eqnarray}
where prime denotes differentiation with respect to $x$. Interestingly, one can
now see that for certain specific forms of the function $f$, the 
nonlinear
ODE (\ref{int03}) becomes the well known and well studied models in the contemporary nonlinear
dynamics literature. For example fixing $f(x)=(\alpha
+\beta x)$, where $\alpha$ and $\beta$ are real arbitrary parameters, either in the NLT or in (\ref{int03}) one can get the generalized modified Emden type equation (MEE)
\begin{eqnarray}  
\ddot{x}+(2\alpha+3\beta
x)\dot{x} +\beta^2x^{3}+2\alpha
\beta x^{2}+(\alpha^2+\lambda)x=0,
\label {int04}
\end{eqnarray}
which has been studied in detail
in Refs. \cite{Hyet,mahomed:1985,leach:1988a,Feix,Dix:1990,Chand2,Chand3,Chand4}. On the other hand the restriction $f(x)=(\alpha+\beta x^2)$ in (\ref{int03}) gives us 
generalized force-free Duffing-van der Pol oscillator (DVP) equation
\begin{eqnarray}      
\ddot{x}+(2\alpha+4\beta
x^2)\dot{x}+\beta^2x^{5} +2\alpha
\beta x^{3}+(\alpha^2
+\lambda)x=0,
\label {int05}
\end{eqnarray}
which has also been studied widely in the literature 
\cite{Chand3,Chand4,Smith:1961,Sawada,Gonz:1983}, 
and so on. Also, in our analysis we have considered a  general nonlocal transformation of the form
\begin{eqnarray}            
U=x^ne^{\int_0^t{(\beta(t') x^m+\alpha(t'))}dt'},
\label {r-nld02}
\end{eqnarray}
where $\alpha(t)$ and
$\beta(t)$ are arbitrary real functions of $t$,
and identified a rather general class of integrable equations of arbitrary order \cite{Chand:07:nld}. We have also developed a method to construct the solutions of the nonlinear oscillators 
(\ref{int04}), (\ref{int05}) and so on.

\subsection{Present work}
Now we turn our attention to the complex case. In this case we take the dependent variable(s) of the linear ODE itself as a complex variable of the independent (real) variable `t'. But the crucial question is how to choose the nonlocal transformation. To begin with, in this work, we complexify the nonlocal transformation (\ref{r-nld02}) in two ways. In the first case (hereafter it is called Type-I) we complexify (\ref{r-nld02}) in the form 
\begin{eqnarray}
z(t)=Z^n e^{\int_0^t(\alpha(t')+\beta(t')|Z|^m)dt'}, \label {sl04}
\end{eqnarray}
where $n$ and $m$ are constants and $\alpha(t)$ and
$\beta(t)$ are arbitrary complex functions of $t$, that is, 
$\alpha(t)=\alpha_{+}(t)+i\alpha_{-}(t)$ and
$\beta(t)=\beta_{+}(t)+i\beta_{-}(t)$. The variable $z(t)$ is the solution of a linear ODE. The motivation for choosing this
form is by direct observation: For instance, choosing $n=1$ and
$\alpha(t)$, $\beta(t)$ as constants in (\ref{sl04}) and substituting the resultant NLT in the first order linear ODE $\dot{z}=0$ we get the following nonlinear ODE, namely 
\begin{eqnarray}
\frac{dZ}{dt}+\alpha Z+\beta |Z|^mZ=0.\label {sl02}
\end{eqnarray}
The choices $m=1$ and $m=2$ in (\ref{sl02}) provide us the Poincar\'e\cite{Kuramoto:84}
and Stuart-Landau\cite{Pikovsky:01} oscillator equations,  respectively.

Now let us consider a NLT of the form
\begin{eqnarray}
z(t)=Z(t) e^{\int_0^t(\alpha+\beta|Z|)dt'}, \label {sl04aa}
\end{eqnarray}
where $\alpha$ and $\beta$ are arbitrary constants and substitute it into the linear second order ODE, $\ddot{z}+\lambda z=0$, where $\lambda$ can be either real or complex. The result is that one gets a complex MEE, 
\begin{eqnarray}
\fl \qquad  \ddot{Z}+\beta Z^2\dot{Z}^{*}+\beta^2
 |Z|^{4}Z
+(2\alpha+3\beta|Z|^2)\dot{Z} 
 +2\alpha \beta Z|Z|^{2}
 +(\alpha^2 +\lambda )Z=0. \label {1-sfeq01a}
\end{eqnarray} 

The restriction $Z$ be real provides us (\ref{int04}). The illustrations given above clearly show that one can identify a class of complex nonlinear oscillator equations of arbitrary order by appropriately choosing the parameters and functions in (\ref{sl04}).

However, recently attempts are also being made to analyse the linearization and integrability properties of certain complexified versions of nonlinear ODEs, which are obtained by assuming the dependent variable to be a complex one in a given differential equation instead of being real or extending it to the complex domain. For example, the well known MEE (\ref{int04}) by a straight forward complexification reads
\begin{eqnarray}
\fl \qquad \ddot{Z}+\beta^2Z^{3}+(2\alpha+3\beta
Z)\dot{Z} +2\alpha
\beta Z^{2}+(\alpha^2
+\lambda)Z=0. \label {1-sfeq01}
\end{eqnarray}

It is clear that now one has to consider a slightly different form of NLT than (\ref{sl04}) in order to deduce this ``straight complexified equations". To identify this type of equations we consider a NLT of the type (hereafter it is called Type-II) 
\begin{eqnarray}
z(t)=Z^ne^{\int_0^t{(\alpha(t')+\beta(t') Z^m)}dt'}, \label {nld02}
\end{eqnarray}
where $n$ and $m$ are real constants and $\alpha(t)$ and $\beta(t)$ are complex functions. To see the outcome let us choose $n=1,\;m=1$ and $\alpha$, $\beta$ as constants and substitute the resultant NLT into the linear ODE, 
$\ddot{z}+\lambda z=0$. In this case one exactly gets the `straight complexified MEE (\ref{1-sfeq01})'.

The above arguments naturally suggest one to conclude that one can explore several complex integrable nonlinear ODEs (corresponding to the real ones) by appropriately choosing the NLT. However, in this article, we confine our attention only on the above two types of NLTs and bring out the resultant integrable complex nonlinear oscillators.

Once the hidden connection between complex linear and nonlinear oscillators is established the question now becomes how to construct the general solution for these two types of complexified nonlinear oscillators, so that complete integrability of the underlying equations can be established and the dynamics of these oscillators understood. Since the nonlinear ODEs are different in each of the above cases one has to adopt a suitable methodology in order to construct the solution. In fact we devise a general procedure for each of the cases and construct the corresponding general solution. The method is same for all orders of the ODEs and depends only on the form of nonlocal transformations. 

We organize the rest of the paper as follows. 
In Sec. 2, we present the
general theory which connects the complex linear ODEs with complex
nonlinear ODEs and the method of finding the general solution. We
consider two types of NLTs to illustrate our theory and obtain
general solution for each of these cases in Sec. 3. In Sec. 4, we explore some of the
physically important complex nonlinear oscillators. We present our
conclusion in Sec. 5.

\section{General theory}
In this section we make use of the main ideas given in the introduction and formulate a general theory which is applicable for arbitrary order and useful for subsequent sections.

Let us consider a $n$th order linear complex ODE of the form
\begin{eqnarray}
\frac{d^nz}{dt^n}+\lambda_1\frac{d^{(n-1)}z}{dt^{(n-1)}}
+\ldots+\lambda_{n-1}\frac{dz}{dt}+\lambda_nz=0, \label
{noe01}
\end{eqnarray}
where $z$ is a complex variable, $\lambda_j$'s, $j=1,2,...n,$ are arbitrary complex 
parameters and $t$ is a real variable. Let us assume that the general solution of
(\ref{noe01}) be  $z(t)=a(t)$, where $a(t)$ is again a complex function. Now let us introduce a nonlocal transformation (NLT) of the form
\begin{eqnarray}
z(t)=g(t,Z)e^{\int_0^t f(t',Z,Z^{*})dt'}, \label {noe02}
\end{eqnarray}
where $f(t,Z,Z^{*})$ and $g(t,Z)$ are arbitrary complex functions of their arguments and $Z$ is the new dependent variable, in (\ref{noe01}) so that the latter becomes a nonlinear ODE of the form
\begin{eqnarray}
\bigg(D_{h}^{(n)}+\lambda_1D_{h}^{(n-1)}+\ldots+\lambda_{n-1}D_{h}^{(1)}
+\lambda_n\bigg)g(t,Z)=0, \label {noe03}
\end{eqnarray}
where $D_{h}^{(n)} = (\frac{d}{d t}+ f(t,Z,Z^{*}))^n$. One may note that even for fixed forms of $f$ and $g$ the resultant nonlinear ODE (\ref{noe03}) turns out to be a very complicated one (see for example equations (\ref{nld03a}) and (\ref{fin1}) for the second order case). The task is to deduce the interesting cases from (\ref{noe03}) which can be integrated explicitly.

Interestingly one can easily show that solving equation (\ref{noe03}) is equivalent to integrating the first order complex nonlinear ODE in $Z$ of the form
\begin{eqnarray}
\dot{Z}=\bigg(\hat{a}(t)-f(t,Z,Z^{*})\bigg)\frac{g(t,Z)}{g_Z(t,Z)}
-\frac{g_{t}(t,Z)}{g_Z(t,Z)}, \quad 
\hat{a}=\frac{\dot{a}}{a}=\hat{a}_{+}+i\hat{a}_{-},\label {noe05}
\end{eqnarray}
where $a(t)$ is the general solution of the linear equation (\ref{noe01}).
This can be derived by using the identity
\begin{eqnarray}
\frac{\dot{z}(t)}{z(t)}=\frac{\dot{a}}{a}=\hat{a}=
\frac{g_Z(t,Z)\dot{Z}(t)
+g_t(t,Z)}{g(t,Z)}+f(t,Z,Z^{*}).
\label {noe04}
\end{eqnarray}
One can note that the equation (\ref{noe05}) is the equivalent first order form of the nonlinear equation (\ref{noe03}). The equivalence between the two equations can be checked by differentiating $(n-1)$ times Eq.(\ref{noe05}) (where $n$ is the order of the equation considered)
 and eliminating $\hat{a}$ in the resulting expression.
 
For certain specific forms of $f$ and $g$, equation (\ref{noe05}) can be
integrated and at least for all these cases, the general solution for the
nonlinear ODE (\ref{noe03}) can be obtained as we explain below in Secs. 3 and 4. In this way one can classify a class of  integrable nonlinear ODEs of any order through NLT of the form (\ref{noe02}).

One may observe that the general solution of (\ref{noe05}) contains 
$2(n+1)$ real integration constants (instead of $2n$), out of which $2n$ constants
are contained in $\hat{a}(t)$ and the remaining two
will appear while integrating (\ref{noe05}) (say $C_{+}$ and $C_{-}$). However, as we demonstrate in the Secs. 3 and 4, out of these $2(n+1)$ constants two of them can be absorbed with the other integration constants and hence the solution
now contains only $2n$ independent integration constants. For example, the general
solution of (\ref{noe01}) can be written in the form
\begin{eqnarray}
\fl \qquad
z(t)=a(t)=(I_{1+}+iI_{1-})e^{(m_{1+}+im_{1-})t}+(I_{2+}+iI_{2-})e^{(m_{2+}+im_{2-})t}+\ldots
\nonumber\\
\qquad\qquad\qquad\qquad\qquad\qquad+(I_{n+}+iI_{n-})e^{(m_{n+}+im_{n-})
t}, \label {nld08}
\end{eqnarray}
where $I_{j+}$'s and $I_{j-}$'s, $j=1,2,\ldots n$, are integration
constants and $m_{j+}$'s and $m_{j-}$'s, $j=1,2,\dots n$, are the
roots of the auxiliary equation of (\ref{noe01}). From (\ref{nld08})
we get
\begin{eqnarray}
\fl \quad
\hat{a}(t)=\frac{\dot{a}}{a}=\frac{\dot{z}}{z}=\frac{m_1(\hat{I}_{1+}+i\hat{I}_{1-})
e^{m_1t} +m_2(\hat{I}_{2+}+i\hat{I}_{2-})e^{m_2t}+\ldots
+m_ne^{m_nt}}
{(\hat{I}_{1+}+i\hat{I}_{1-})e^{m_1t}+(\hat{I}_{2+}+i\hat{I}_{2-})e^{m_2t}+\ldots
+e^{m_nt}},\label {nld10}
\end{eqnarray}
where $\hat{I}_{j\pm}=\frac{(I_{j\pm}I_{n+}\pm
I_{j\mp}I_{n-})}{(I_{n+}^2+I_{n-}^2)}$, $j=1,2,\ldots n-1$ and
$m_j=m_{j+}+im_{j-}$, $j=1,2,\ldots n$. Note that equation
(\ref{nld10}) has only $2(n-1)$ arbitrary constants, so that the
general solution (\ref{noe05}) contains $2n$ arbitrary constants,
namely $\hat{I}_j=\hat{I}_{j+}+i\hat{I}_{j-}$, $j=1,2,\ldots n-1$,
$C_{+}$ and $C_{-}$.

In the following section, we consider two important forms of NLTs as mentioned in the introduction and discuss the methods of constructing the general solutions of the associated nonlinear ODEs.

\section{Nonlocal transformation of type-I}
Let us consider a NLT of the form (\ref{sl04}) and substitute it into the linear ODE (\ref{noe01}). As a result one gets an $n$th order nonlinear ODE in the new variable $Z$. The solution
for this nonlinear ODE can be obtained by solving the equivalent first order ODE (vide equation (\ref{noe05}))
\begin{eqnarray}
\dot{Z}=\bigg(\hat{a}(t)-\alpha(t)\bigg)\frac{Z}{n}
-\frac{\beta(t)}{n}|Z|^{m}Z, \label {sl07}
\end{eqnarray}
where $\hat{a}=\dot{z}/z$. 


To integrate (\ref{sl07}) we introduce polar coordinates $Z=Re^{i\phi}$ so that equation (\ref{sl07}) can be split into two real equations of the form
\begin{eqnarray}
\dot{R}=\frac{1}{n}(\hat{a}_{+}-\alpha_{+} -\beta_{+}R^{m})R,\quad
\dot{\phi}=\frac{1}{n}(\hat{a}_{-}-\alpha_{-}-\beta_{-}R^m), \label
{sl07a}
\end{eqnarray}
where we assumed
$\hat{a}(t)=\frac{\dot{a}}{a}=\hat{a}_{+}(t)+i\hat{a}_{-}(t)$. One may observe that the $R$ equation does not contain the phase variable $\phi$ and turns out to be the Bernoulli equation. As a result the general solution can be immediately found. Substituting this form of $R$ into the $\phi$ equation and integrating the resultant ODE one can obtain the form of $\phi$. Once $R$ and $\phi$ are known the general solution of equation (\ref{sl07}) can be fixed in the form 
\begin{eqnarray}
Z=Re^{i\phi}=R(\cos(\phi)+i\sin(\phi)),\label {sl08a}
\end{eqnarray}
where
\begin{eqnarray}
\fl \qquad
R(t)=e^{-\frac{1}{n}\int_0^t(\alpha_{+}(t')-\hat{a}_{+}(t'))dt'}\bigg[C_{+}
+\frac{m}{n}\int_0^t\beta_{+}(t')e^{-\frac{m}{n}\int_0^{t'}(\alpha_{+}(t'')-\hat{a}_{+}(t''))dt''}dt'
\bigg]^{\frac{-1}{m}},
\nonumber\\
\fl \qquad \phi(t)=
C_{-}-\frac{1}{n}\int_0^t\bigg(\alpha_{-}(t')-\hat{a}_{-}(t')-\beta_{-}(t')R(t')^m\bigg)dt',\label
{sl08}
\end{eqnarray}
Here $C_+$ and $C_-$ are integration constants.\\
We note here that the solution (\ref{sl08}) can be written more elegantly by using the identities $r(t)=e^{\int \hat{a}_{+} dt}$ and $\theta(t)=\int \hat{a}_{-} dt$. These two identities can be derived easily as follows.

We split the complex solution $\hat{a}(t)$ in the form 
$\hat{a}(t)=\frac{\dot{a}}{a}=\hat{a}_{+}(t)+i\hat{a}_{-}(t)$. In polar form, $z=re^{i\theta}$, this expression is equivalent to 
$\hat{a}(t)=\frac{\dot{z}}{z}=\frac{\dot{r}}{r}+i\dot{\theta}$.
Equating the right hand sides of these two equations, one arrives at the conclusion
\begin{eqnarray}
r(t)=e^{\int \hat{a}_{+} dt},
\quad \theta(t)=\int \hat{a}_{-} dt,
\label{sl08ab}
\end{eqnarray}
where the integration constants will be subsumed with other integration constants.
As a consequence the solution (\ref{sl08}) can be written in a more compact form as
\begin{eqnarray}
\fl \qquad
R(t)=r(t)^{\frac{1}{n}}e^{-\frac{1}{n}\int_0^t\alpha_{+}(t')tdt'}\bigg[C_{+}
+\frac{m}{n}\int_0^t r(t')^{\frac{m}{n}}\beta_{+}(t')e^{-\frac{m}{n}\int_0^{t'}\alpha_{+}(t'')dt''}dt'
\bigg]^{\frac{-1}{m}},
\nonumber\\
\fl \qquad \phi(t)=
C_{-}+\frac{\theta(t)}{n}-\frac{1}{n}\int_0^t\bigg(\alpha_{-}(t')-\beta_{-}(t')R(t')^m\bigg)dt'.\label
{sl08ac}
\end{eqnarray}
From the computational point of view (\ref{sl08ab}) is more useful 
as we illustrate in the following.

\subsection{Second order ODEs}
By using the nonlocal transformation (\ref{sl04}) we identify the nonlinear oscillator equation of second order that is connected with the complex harmonic oscillator equation
\begin{eqnarray}
\ddot{z}+\lambda z=0, \label {nld01}
\end{eqnarray}
where $\lambda$ is a real parameter (it can be complex too). To illustrate the main idea we also fix the arbitrary parameters in the NLT (\ref{sl04}) in the following way, that is, $n=1,\;m=q,\;\alpha(t)=\alpha_{+}+i\alpha_{-}$ and $\beta(t)=\beta_{+}+i\beta_{-}$, where $\alpha_{\pm}$ and $\beta_{\pm}$ are real constants, so that the NLT now becomes
\begin{eqnarray}
z(t)=Z(t) e^{\int_0^t(\alpha+\beta |Z|^q)dt'}. \label {nld01aa}
\end{eqnarray}
The NLT (\ref{nld01aa}) transforms the linear equation (\ref{nld01}) to the nonlinear form 
\begin{eqnarray}
\ddot{Z}+\frac{q}{2}\beta|Z|^{q-2}Z^2\dot{Z}^{*}+\beta^2
 |Z|^{2q}Z
+(2\alpha+(2+\frac{q}{2})\beta|Z|^q)\dot{Z}
\nonumber\\\qquad \qquad \qquad \qquad
 +2\alpha \beta Z|Z|^{q}
 +(\alpha^2+\lambda )Z=0. \label {sfeq01a}
\end{eqnarray}

Now we construct the general solution of the nonlinear oscillator equation from the solution of (\ref{nld01}) by using the idea given in previous section.
The general solution of (\ref{nld01}) can be written as 
\begin{eqnarray}
z(t)=a(t)=(I_{1}+iI_{2})\sin\omega t+(I_{3}+iI_{4})\cos\omega t, \label
{cal02}
\end{eqnarray}
where $\omega=\sqrt{\lambda}$ and $I_i$, $i=1,...4,$ are integration constants. 
From (\ref{cal02}) we can fix $\hat{a}(t)$ in the form
\begin{eqnarray}
\hat{a}(t)&=&\frac{\dot{z}}{z}=\frac{\dot{a}}{a}=\frac{\omega((I_{1}+iI_{2})\cos\omega
t-(I_{3}+iI_{4})\sin\omega
t)}{(I_{1}+iI_{2})\sin\omega t+(I_{3}+iI_{4})\cos\omega t}\nonumber\\
&=&\frac{\omega((A+iB)\cos\omega t-\sin\omega t)}{(A+iB)\sin\omega
t+\cos\omega t},\label {cal03}
\end{eqnarray}
where $A=\frac{(I_{1}I_{3}+
I_{2}I_{4})}{(I_{3}^2+I_{4}^2)}$ and $B=\frac{(I_{2}I_{3}-
I_{1}I_{4})}{(I_{3}^2+I_{4}^2)}$. Separating the real and imaginary parts, we get
\begin{eqnarray}
\fl \quad \hat{a}(t)&=&\frac{\omega[(A\cos\omega t-\sin\omega t)(A\sin\omega
t+\cos\omega t)+B^2\cos\omega t\sin\omega t+iB]}{(A\sin\omega
t+\cos\omega t)^2+B^2\sin^2\omega t}\label {cal03aa}
\end{eqnarray}
from which one finds
\begin{eqnarray}
\fl \qquad \hat{a}_{+}(t)&=&\frac{\omega((A\cos\omega t-\sin\omega t)(A\sin\omega
t+\cos\omega t)+B^2\cos\omega t\sin\omega t}{(A\sin\omega
t+\cos\omega t)^2+B^2\sin^2\omega t},\nonumber\\
\fl \qquad \hat{a}_{-}(t)&=&\frac{B\omega}{(A\sin\omega
t+\cos\omega t)^2+B^2\sin^2\omega t}. \label{cal04}
\end{eqnarray}
The quantities $r(t)$ and $\theta(t)$ in (\ref{sl08ac}) can be obtained from (\ref{cal04}) as
\begin{eqnarray}
 r(t)&=&e^{\int_0^t\hat{a}_{+}(t')dt'}=\bigg((A\sin\omega t+\cos\omega
t)^2+B^2\sin^2\omega t\bigg)^{\frac{1}{2}},\nonumber\\
\theta(t)&=&\int_0^t\hat{a}_{-}(t')dt'=tan^{-1}\bigg(\frac{B\sin\omega t}{(A\sin\omega
t+\cos\omega t)}\bigg).
 \label {cal06}
\end{eqnarray}
The general solution of equation (\ref{sfeq01a}) can be obtained easily from
(\ref{cal06}) and (\ref{sl08ac}) as
\begin{eqnarray}
R(t)&=&r(t)e^{-\alpha_{+}t}\bigg[C_{+}
+q\beta_{+}\int_0^t r(t')^qe^{-q\alpha_{+}t'}dt'
\bigg]^{\frac{-1}{q}},
\nonumber\\
\phi(t)&=&
C_{-}+\theta(t)-\alpha_{-}t
+\frac{\beta_{-}}{q\beta_{+}}\log\bigg[C_{+}
+q\beta_{+}\int_0^t r(t')^qe^{-q\alpha_{+}t'}dt'
\bigg].\label
{sol-1}
\end{eqnarray}

Equations~(\ref{sfeq01a}) is indeed a complex version of several physically important models. For
example, the choice $q=1$ in (\ref{sfeq01a}) gives us the generalized complex modified
Emden type equation (CMEE) of the form (\ref{1-sfeq01a}).
The general solution of the equation can be deduced easily from (\ref{sol-1}) with the same restriction, and $q=1$. For simplicity let us consider the case $q=1$ and $\alpha=0$. Then the solution reads as $Z=Re^{i\phi}$, where $R$ and $\phi$ are given by
\begin{eqnarray}
R(t)&=&\frac{\bigg((A\sin\omega t+\cos\omega t)^2+B^2\sin^2\omega t\bigg)^{\frac{1}{2}}}{\bigg[C_{+}
+\beta_{+}\int_0^t \bigg((A\sin\omega t'+\cos\omega t')^2+B^2\sin^2\omega t'\bigg)^{\frac{1}{2}}dt'
\bigg]},
\nonumber\\
\phi(t)&=&
C_{-}+tan^{-1}\bigg(\frac{B\sin\omega t}{(A\sin\omega t+\cos\omega t)}\bigg)
\nonumber\\
\fl \qquad\qquad \qquad &&\hspace{-0.5cm}
+\frac{\beta_{-}}{\beta_{+}}\log\bigg[C_{+}
+\beta_{+}\int_0^t \bigg((A\sin\omega t'+\cos\omega t')^2+B^2\sin^2\omega t'\bigg)^{\frac{1}{2}}dt'
\bigg].\label
{sol-1a}
\end{eqnarray}


For the choice $q=2$ and $\alpha=0$, Eq.~(\ref{sfeq01a}) becomes the generalized complex Duffing-van der Pol (CDVP)
oscillator equation of the form
\begin{eqnarray}
\ddot{Z}+\beta Z^2\dot{Z}^{*}+\beta^2
 |Z|^{4}Z
+3\beta|Z|^2\dot{Z}+\lambda Z=0, \label {2-sfeq01a}
\end{eqnarray}
The general solution of (\ref{2-sfeq01a}) can be written down from (\ref{sol-1}) by choosing the values $q=2$ and $\alpha=0$ on the right hand side expressions. The solution is given by
\begin{eqnarray}
R(t)&=&r(t)\bigg(C_{+}-\frac{\beta_{+}}{2\omega}(4A\cos[\omega t]+(A^2+B^2-1)\sin[2\omega t]
\nonumber\\
\fl \qquad\qquad &&\qquad\qquad-2(A^2+B^2+1) \omega t)\bigg)^{\frac{-1}{2}},
\nonumber\\
\phi(t)&=&
C_{-}+\theta(t)
+\frac{\beta_{-}}{2\beta_{+}}\log\bigg[C_{+}-\frac{\beta_{+}}{2\omega}(4A\cos[\omega t]\nonumber\\\fl \qquad\qquad &&+(A^2+B^2-1)\sin[2\omega t]-2(A^2+B^2+1) \omega t)
\bigg],\label
{sol-1b}
\end{eqnarray}
where $r(t)$ and $\theta(t)$ are given by Eqs.(\ref{cal06}).\\
Now we choose $q=-2$ and $\alpha=0$ in equation (\ref{sfeq01a}) so that the latter becomes
\begin{eqnarray}
 \ddot{Z}+(\beta Z
 -Z^2\dot{Z}^{*}+|Z|^{2}\dot{Z})\beta |Z|^{-4}
 +\lambda Z=0, \label {3-sfeq01a}
\end{eqnarray}
The general solution of the equation can be fixed easily from (\ref{sol-1}) by taking $q=-2$ and $\alpha=0$ respectively. The resultant solution reads as
\begin{eqnarray}
R(t)&=&r(t)\bigg(\bigg(C_{+}-\frac{2\beta_{+}}{\omega B}\tan^{-1}\bigg[\frac{A+(A^2+B^2)\tan[\omega t]}{B}\bigg]\bigg)
\bigg)^{\frac{1}{2}},\nonumber\\
\phi(t)&=&C_{-}+\theta(t)\nonumber\\
&-&\frac{\beta_{-}}{2\beta_{+}}\log\bigg[\bigg(C_{+}-\frac{2\beta_{+}}{\omega B}\tan^{-1}\bigg[\frac{A+(A^2+B^2)\tan[\omega t]}{B}\bigg]\bigg)
\bigg].\label
{sol-1c}
\end{eqnarray}
The associated real equation reads as
\begin{eqnarray}
\ddot{X}+\beta^2X^{-3}+\lambda X=0, \label {3-sfeq01b}
\end{eqnarray}
namely the Pinney equation which is another important nonlinear ODE that arises in different
areas of physics and has been studied in detail in Ref.
\cite{Pinney,Lewis}.\\
In a similar way one can consider the general case $\ddot{z}+\lambda_1 \dot{z}+\lambda_2 z=0$ and identify the more general nonlinear oscillator equation, namely
\begin{eqnarray}
\ddot{Z}+(n-1) \frac{\dot{Z}^2}{Z}
+\frac{m}{2n}\beta|Z|^{m-2}Z^2\dot{Z}^{*}+\bigg(c_1(t)+(2+\frac{m}{2n})\beta|Z|^m\bigg)\dot{Z}
\nonumber\\ \qquad \qquad \qquad\qquad\qquad\quad
 +\bigg(c_2(t)+\frac{\beta^2}{n} |Z|^{m}
\bigg)Z|Z|^{m}+c_3(t)Z=0, \label {nld03a}
\end{eqnarray}
where
\begin{eqnarray}
 c_1(t)&=&\bigg(2\alpha(t)+\lambda_1\bigg),\;\;
c_2(t)=\frac{1}{n}\bigg( \dot{\beta(t)}+2\alpha(t) \beta(t) +\lambda_1\beta(t)
\bigg), \nonumber\\ c_3(t)&=&\frac{1}{n}\bigg(\dot{\alpha(t)}+\alpha(t)^2+\alpha(t)
\lambda_1 +\lambda_2 \bigg),
 \label {nld04}
\end{eqnarray}
by the nonlocal transformation (\ref{sl04}). From the above discussed procedure, we can find the general solution of the complex nonautonomous nonlinear ODE (\ref{nld03a}) using the general solution of the linear ODE straightforwardly.

\section{Nonlocal transformation of type-II}
Next, we consider the NLT of the form (\ref {nld02}). Again the action of this NLT on the linear ODE (\ref{noe01}) gives us another class of nonlinear ODEs. We construct the solution of these nonlinear ODEs  by considering the equivalent first order nonlinear ODE  of the form (which is of course different from Eq.(\ref{sl07}))
\begin{eqnarray}
\dot{Z}=\bigg(\hat{a}(t)-\alpha(t)\bigg)\frac{Z}{n}
-\frac{\beta(t)}{n}Z^{m+1}, \label {nld06a}
\end{eqnarray}
where $\hat{a}=\frac{\dot{a}}{a}=\hat{a}_{+}+i\hat{a}_{-}$, which can be obtained simply by substituting (\ref{nld02}) into equation (\ref{noe05}). 

To integrate equation (\ref{nld06a}) we introduce a transformation
$Z=U^{-\frac{1}{m}}$, where $U$ is a complex variable,  in (\ref{nld06a}) so that the latter becomes a linear ODE of the form
\begin{eqnarray}
\dot{U}=m\bigg(\alpha(t)-\hat{a}(t)\bigg)\frac{U}{n}
+m\frac{\beta(t)}{n}. \label {nld06}
\end{eqnarray}
Now we split this complex linear equation (\ref{nld06}) into two real equations by taking $U=U_{+}+iU_{-}$, with $\alpha(t)=\alpha_{+}+i\alpha_{-},\; \hat{a}(t)=\hat{a}_{+}+i\hat{a}_{-}$, and obtain
\begin{eqnarray}
\dot{U}_{\pm}=\gamma_1 U_{\pm}\mp\gamma_2
U_{\mp}+\frac{m}{n}\beta_{\pm}, \label {nld06b}
\end{eqnarray}
where $\gamma_1=\frac{m}{n}(\alpha_{+}-\hat{a}_{+})$
and $\gamma_2=\frac{m}{n}(\alpha_{-}-\hat{a}_{-})$. Solving equation
(\ref{nld06b}) one finds that
\begin{eqnarray}
U_{\pm}(t)=e^{\int_0^t\gamma_1dt'}\bigg(A_{\pm}(t)\cos[\int_0^t\gamma_2dt']\mp
A_{\mp}(t)\sin[\int_0^t\gamma_2dt']\bigg), \label {nld06c}
\end{eqnarray}
with
\begin{eqnarray}
\fl \qquad \quad A_{\pm}(t)=C_{\pm}+\frac{m}{n}\int_0^t
e^{-\int_0^t\gamma_1dt'}\bigg(\beta_\pm\cos[\int_0^t\gamma_2dt']\pm
\beta_\mp\sin[\int_0^t\gamma_2dt']\bigg)dt', \label {nld06d}
\end{eqnarray}
where $C_{\pm}$ are the two integration constants. 

Making use of the transformation $U=Z^{-m}$ one can 
obtain the real and complex parts of the general solution for equation 
(\ref{nld06a}) in the form
\begin{eqnarray}
\fl \;\;
Z_{+}=(U_{+}^2+U_{-}^2)^{\frac{-1}{2m}}\cos\bigg[\frac{1}{m}tan^{-1}(\frac{U_{-}}{U_{+}})\bigg],\;\;
Z_{-}=-(U_{+}^2+U_{-}^2)^{\frac{-1}{2m}}\sin\bigg[\frac{1}{m}tan^{-1}(\frac{U_{-}}{U_{+}})\bigg].
\label {nld07}
\end{eqnarray}
from which the general solution of the nonlinear ODE can be written as
\begin{eqnarray}
\fl \quad
R(t)&=&r(t)^{\frac{1}{n}}e^{-\frac{1}{n}\int_0^t\alpha_{+}dt'}\bigg(A_{+}^2+A_{-}^2\bigg)^{\frac{-1}{2m}},\nonumber\\
\fl \quad 
\phi(t)&=&-\frac{1}{m}tan^{-1}\bigg[\frac{A_{-}\cos[\frac{m}{n}(\theta(t)-\int_0^t\alpha_{-}dt')]+
A_{+}\sin[\frac{m}{n}(\theta(t)-\int_0^t\alpha_{-}dt')]}{A_{+}\cos[\frac{m}{n}(\theta(t)-\int_0^t\alpha_{-}dt')]-
A_{-}\sin[\frac{m}{n}(\theta(t)-\int_0^t\alpha_{-}dt')]}\bigg],
\label {nld07aa}
\end{eqnarray}
where
\begin{eqnarray}
\fl \quad  A_{\pm}(t)=C_{\pm}+\frac{m}{n}\int_0^t
 r(t')^{\frac{m}{n}}e^{-\frac{m}{n}\int_0^{t'}\alpha_{+}dt''}
\bigg(\beta_ \pm\cos[\frac{m}{n}(\theta(t')-\int_0^{t'}\alpha_{-}dt'')]\nonumber\\
\qquad \qquad \pm
\beta_ \mp\sin[\frac{m}{n}(\theta(t')-\int_0^{t'}\alpha_{-}dt'')]\bigg)dt'.
 \label {nld07ab}
\end{eqnarray}

From the above equation, we can discuss the nature of the solution. For convenience, first ‪let us consider $m$ in a fractional form, that is $m=\frac{p} {q}$, and assume $p=1$ and $q=$integer. From the transformation $U=Z^{-m}$, the solution becomes $Z=U^{\frac{-q} {p}}$ and we will get only one solution which is singlevalued. But for negative integer values of $m$ except $-1$, we can rewrite the solution as $Z^{m}=U$. So it leads to $m$ roots, thereby leading to a multivaluedness of the solution. These concepts are briefly explained in the following section. 

\subsection{Second order ODEs}
Consider the nonlocal transformation (\ref{nld02}) with the parametric choices $n=1,\;\alpha(t)=\alpha_{+}+i\alpha_{-}$ and $\beta(t)=\beta_{+}+i\beta_{-}$, where $\alpha_{\pm}$ and $\beta_{\pm}$ are real constants. Then the NLT becomes
\begin{eqnarray}
z(t)=Z(t) e^{\int_0^t(\alpha+\beta Z^m)dt'} \label {nld01ab}.
\end{eqnarray}
Now the NLT (\ref{nld01ab}) transforms the linear ODE (\ref{nld01}) into the nonlinear ODE, 
\begin{eqnarray}
\ddot{Z}+\beta^2Z^{2m+1}+(2\alpha+(2+m)\beta
Z^m)\dot{Z} +2\alpha
\beta Z^{m+1}+(\alpha^2
+\lambda)Z=0.\label {sfeq01}
\end{eqnarray}
For particular choices of $m$, Eq.(\ref{sfeq01}) will give physically important dynamical systems. Singlevalueness or multivalueness of the solution may result in depending on the choice of $m$. First, we discuss the cases where the solutions are single valued.
\subsubsection{Case 1-Singlevalued solution}
For the choice, $m=\frac{p} {q}$, $p=1$ and $q=$ positive integer, the solution is always singlevalued. To explain this nature, we choose different values for $q$ as in the following:

\begin{enumerate}
\item For the choice $q=1$, Eq.(\ref{sfeq01}) gives the generalized complex modified
Emden type equation (CMEE) (\ref{1-sfeq01}). Further with the choice $\alpha =0$ in (\ref{1-sfeq01}) the solution $Z=Re^{i\phi}$ can be expressed as 
\begin{equation}\hspace{-2cm} \nonumber
Z_+=R\cos(\phi)=\frac{\omega B_-\bigg((A\sin\omega t+\cos\omega t)^2+B^2\sin^2\omega t\bigg)^{\frac{1}{2}}}{\sqrt{(A_{+}^2+A_{-}^2)(B_+^2+B_-^2)}},\label{hl1} 
\end{equation}
\begin{equation}
\hspace{-2cm}
Z_-=R\sin(\phi)=-\frac{\omega B_+\bigg((A\sin\omega t+\cos\omega t)^2+B^2\sin^2\omega t\bigg)^{\frac{1}{2}}}{\sqrt{(A_{+}^2+A_{-}^2)(B_+^2+B_-^2)}}\label{hl2}
\end{equation}
with
\begin{eqnarray}
\fl \qquad  A_{\pm}(t)=(C_{\pm}\omega-(A\beta_{\pm}\pm B\beta_{\mp})\cos[\omega t]+\beta_{\pm}\sin[\omega t])\nonumber\\
\fl \qquad  B_{\pm}(t)=(A^2+B^2-1)\beta_{\pm}\sin[2\omega t-2(AC_{\pm}\mp B C_{\mp})\sin[\omega t]\nonumber\\
\qquad\qquad\qquad
+2A\beta_{\pm}\cos[2\omega t]-2C_{\pm}\omega \cos[\omega t]\pm2B\beta_{\mp}.\label{hl3}
\end{eqnarray}

Note that the real valued case of system (\ref{sfeq01}) with $q=1$ has been shown recently \cite{chi1} to be an exactly quantizable PT-symmetric non-Hermitian Hamiltonian system. As noted in the Introduction, complex versions of such PT-symmetric systems are equally important. In this connection, the above complex solution (\ref{hl1})-(\ref{hl3}) will have important consequence.
\item For $q=2$ , the solution of (\ref{sfeq01}) becomes
\begin{eqnarray}
\fl \quad
R(t)&=&r(t)\bigg(A_{+}^2+A_{-}^2\bigg)^{-1},\nonumber\\
\fl \quad 
\phi(t)&=&-2tan^{-1}\bigg[\frac{A_{-}\cos[\frac{1}{2}(\theta(t))]+
A_{+}\sin[\frac{1}{2}(\theta(t))]}{A_{+}\cos[\frac{1}{2}(\theta(t))]-
A_{-}\sin[\frac{1}{2}(\theta(t))]}\bigg],
\label {nld07aa1}
\end{eqnarray}
where
\begin{eqnarray}
\fl \quad  A_{\pm}(t)=C_{\pm}+\frac{1}{2}\int_0^t
 r(t')^{\frac{1}{2}}\bigg(\beta_ \pm\cos[\frac{1}{2}(\theta(t'))]\pm
\beta_ \mp\sin[\frac{1}{2}(\theta(t'))]\bigg)dt'.
 \label {nld07ab12}
\end{eqnarray}
\item For $q=3$, the solution of (\ref{sfeq01}) becomes
\begin{eqnarray}
\fl \quad
R(t)&=&r(t)\bigg(A_{+}^2+A_{-}^2\bigg)^{-\frac{3} {2}},\nonumber\\
\fl \quad 
\phi(t)&=&-3tan^{-1}\bigg[\frac{A_{-}\cos[\frac{1}{3}(\theta(t))]+
A_{+}\sin[\frac{1}{3}(\theta(t))]}{A_{+}\cos[\frac{1}{3}(\theta(t))]-
A_{-}\sin[\frac{1}{3}(\theta(t))]}\bigg],
\label {nld07aa1}
\end{eqnarray}
where
\begin{eqnarray}
\fl \quad  A_{\pm}(t)=C_{\pm}+\frac{1}{3}\int_0^t
 r(t')^{\frac{1}{3}}
\bigg(\beta_ \pm\cos[\frac{1}{3}(\theta(t'))]\pm
\beta_ \mp\sin[\frac{1}{3}(\theta(t'))]\bigg)dt'.
 \label {nld07ab12}
\end{eqnarray}
\end{enumerate}
Here $r(t)$ and $\theta(t)$ are given in Eq.(\ref{cal06}). 
In the following subsection, we will discuss the cases where the solution becomes multi-valued.

\subsubsection{Multivalueness of the solution}
Let us consider the case $m=-2$ and $\alpha=0$, then equation (\ref{sfeq01}) becomes 
\begin{eqnarray}
\ddot{Z}+\beta^2Z^{-3}+\lambda Z=0.\label{p1}
\end{eqnarray}
From the transformation $Z^{-m}=U$, we can get two roots. So it leads to two solutions which are given as follows:
The first solution $Z_1$ is given by
\begin{eqnarray}
\fl \;\;
Z_{+}=(U_{+}^2+U_{-}^2)^{\frac{1}{4}}\cos\bigg[\frac{1}{2}tan^{-1}(\frac{U_{-}}{U_{+}})\bigg],\;\;
Z_{-}=(U_{+}^2+U_{-}^2)^{\frac{1}{4}}\sin\bigg[\frac{1}{2}tan^{-1}(\frac{U_{-}}{U_{+}})\bigg].
\label{soln1}
\end{eqnarray}
The second solution $Z_2$ is given by
\begin{eqnarray}
\fl \;\;
Z_{+}=-(U_{+}^2+U_{-}^2)^{\frac{1}{4}}\cos\bigg[\frac{1}{2}tan^{-1}(\frac{U_{-}}{U_{+}})\bigg],\;\;
Z_{-}=-(U_{+}^2+U_{-}^2)^{\frac{1}{4}}\sin\bigg[\frac{1}{2}tan^{-1}(\frac{U_{-}}{U_{+}})\bigg].
\label{soln2}
\end{eqnarray}
where $r(t),\;\theta(t),\;A_+$ and $A_-$ are given by Eqs.(\ref{cal06}) and (\ref{nld07ab}). The associated real equation reads the same as Eq.(\ref{3-sfeq01b}).
For $m=-3$, equation (\ref{sfeq01}) becomes 
\begin{eqnarray}
\ddot{Z}+\beta^2 Z^{-5}-\beta Z^{-3}\dot{Z}+\lambda Z=0.\label{g1}
\end{eqnarray}
We will get three roots in this case. So we can get three solutions leading to multivalueness of the solution. 
The real equation associated with Eq.(\ref{g1}) reads as 
\begin{equation}
\ddot{X}+\beta^2 X^{-5}-\beta X^{-3}+\lambda X=0.
\end{equation}

Finally, we can consider the more general linear ODE $\ddot{z}+\lambda_1\dot{z}+\lambda_2 z=0$ for which the nonlinear ODE becomes 
\begin{eqnarray}
\ddot{Z}+(n-1)\frac{\dot{Z}^2} {Z}+\frac{\beta^2} {n}Z^{2m+1}+(c_1(t)+(2+\frac{m} {n})\beta Z^m)\dot{Z}\nonumber \\
\quad\quad\quad\quad\hspace{4cm}+c_2(t)Z^{m+1}+c_3(t)Z=0, \label{fin1}
\end{eqnarray}
where the coefficients are given in Eq.(\ref{nld04}), through the nonlocal transformation (\ref{nld02}). From the above discussed procedure, we can find the solution of the nonlinear equation (\ref{fin1}) from the general solution of the linear ODE systematically.

\section{Third order ODEs}
Let us next consider a linear third order ODE of the form
\begin{eqnarray}
\tdot{z}+\lambda_1 \ddot{z}+\lambda_2 \dot{z}+\lambda_3z=0, \label
{toe01}
\end{eqnarray}
where $\lambda_1,\;\lambda_2$ and $\lambda_3$ are arbitrary
constants. The nonlocal transformations (\ref{sl04}) and
(\ref{nld02}) transform (\ref{toe01}) to the nonlinear ODEs of the
forms
\begin{eqnarray}
\fl (I):\quad
\tdot{Z}+[3(n-1)\frac{\dot{Z}}{Z}+d_1(t,Z,Z^{*})]\ddot{Z}
+\frac{m}{2n}\beta|Z|^{m-2}Z^2\ddot{Z}^{*}
+(n-1)(n-2)\frac{\dot{Z}^3}{Z^2}\nonumber\\
\fl
\qquad\qquad+(\frac{m}{2n}(3n+m)\beta|Z|^{m-2}Z\dot{Z^{*}}+d_2(t,Z,Z^{*})\dot{Z}
+d_3(t,Z,Z^{*}))\dot{Z}\nonumber\\
\fl \qquad \qquad\qquad\quad
+(\frac{m}{4n}(m-2)\beta|Z|^{m-4}Z^3\dot{Z^{*}}+d_4(t,Z,Z^{*}))\dot{Z^{*}}
+\frac{\beta^3}{n}|Z|^{3m}Z\nonumber\\
\fl \qquad\qquad\qquad\qquad\qquad\qquad\qquad\qquad+d_8(t)|Z|^{2m}Z
+d_9(t)|Z|^{m}Z+d_{10}(t)Z=0,
\label {toe02a} \\
\fl (II):\quad \tdot{Z}+[3(n-1)\frac{\dot{Z}}{Z}+d_5(t,Z)]\ddot{Z}
+(n-1)(n-2)\frac{\dot{Z}^3}{Z^2}+d_6(t,Z)\dot{Z}^2 \nonumber\\
\fl \qquad\qquad\qquad
+d_{7}(t,Z)\dot{Z}+\frac{\beta^3}{n}Z^{3m+1}+d_8(t)Z^{2m+1}+d_9(t)Z^{m+1}+d_{10}(t)Z=0,
\label {toe02}
\end{eqnarray}
respectively, where
\begin{eqnarray}
\fl \quad d_1(t,Z,Z^{*})=3\alpha +(3
+\frac{m}{2n})\beta|Z|^{m}+\lambda_1, \nonumber\\\fl \quad
d_2(t,Z,Z^{*})=\bigg(\frac{(n-1)}{Z}(3(\alpha+\beta
|Z|^{m})+\lambda_1)+\frac{m}{4n}\beta|Z|^{m-2}Z^{*}(6n +m-2)\bigg),
\nonumber\\\fl \quad
d_3(t,Z,Z^{*})=\bigg(3(\dot{\alpha}+\dot{\beta}|Z|^m+(\alpha+\beta|z|^m)^2)+\frac{1}{n}((\alpha
+\beta|z|^m)(2n\lambda_1+\frac{m}{2}\beta|Z|^m)\nonumber\\
\fl \qquad\qquad\qquad+m\beta^2|Z|^{2m}+m\alpha\beta|Z|^m
+\frac{m}{2}\beta\lambda_1|Z|^m+\lambda_2+m\dot{\beta}|Z|^m)\bigg),
\nonumber\\\fl \quad
d_4(t,Z,Z^{*})=\bigg(m|Z|^{m-2}Z^2(\frac{\beta}{2n}(\alpha+\beta|z|^m)
+\frac{\beta^2}{n}|Z|^m+\frac{\dot{\beta}}{n}+\frac{\alpha\beta}{n}
+\frac{\lambda_1\beta}{2n})\bigg)\nonumber\\
\fl \quad
d_5(t,Z)=\frac{1}{n}\bigg(3n\alpha+n\lambda_1+\beta(m+3n)Z^m\bigg),
\nonumber\\\fl \quad
d_6(t,Z)=\frac{1}{n}\bigg(\bigg(m(m+2n)+(n-1)(m+3n)\bigg)\beta
Z^{m-1} +n(n-1)(3\alpha+c_3)Z^{-1}\bigg), \nonumber\\\fl \quad
d_{7}(t,Z)=\frac{1}{n}\bigg(\bigg(3\alpha\beta(m+2n)+\dot{\beta}(2m+3n)
+\lambda_1(m+2n)\beta\bigg)Z^{m} \nonumber\\\qquad\qquad
+3\beta^2(m+n)Z^{2m}+n\bigg(3\dot{\alpha}+3\alpha^2+2\lambda_1\alpha+
\lambda_2\bigg)\bigg),
\nonumber\\
\fl \quad d_8(t)=\frac{1}{n}\bigg(3\beta\dot{\beta}+3\alpha\beta^2
+\beta^2\lambda_1\bigg),
\nonumber\\
\fl \quad
d_9(t)=\frac{1}{n}\bigg(\ddot{\beta}+3(\dot{\alpha}\beta+\alpha\dot{\beta})
+3\beta\alpha^2+(\lambda_1\dot{\beta}+\lambda_2\beta)+2\lambda_1\alpha
\beta\bigg), \nonumber\\\fl \quad
d_{10}(t)=\frac{1}{n}\bigg(\ddot{\alpha}+3\alpha\dot{\alpha}+\alpha^3
+\lambda_1(\dot{\alpha}+\alpha^2)+\alpha \lambda_2+\lambda_3\bigg).
\label {toe03}
\end{eqnarray}

For the parametric choice $n=1,\;m=1,\;\alpha(t)=0$ and
$\beta(t)=\beta$ Eqs. (\ref{toe02a}) and (\ref{toe02}) are nothing
but a complex generalized special cases of Chazy equation XII, that is,
\begin{eqnarray}
\fl (I):\quad \tdot{Z}+(\lambda_1+\frac{7}{2}\beta|Z|)\ddot{Z}
+\frac{1}{2}\beta|Z|^{-1}Z^2\ddot{Z}^{*}
+2\beta|Z|^{-1}Z\dot{Z^{*}}\dot{Z}+\frac{5}{4}\beta|Z|^{-1}Z^{*}\dot{Z}^2
\nonumber\\
+(\frac{9}{2}\beta^2|Z|^2
+\frac{5}{2}\beta\lambda_1|Z|+\lambda_2)\dot{Z}
+(\frac{3}{2}\beta^2|Z|+
\frac{\lambda_1\beta}{2})|Z|^{-1}Z^2\dot{Z^{*}}
\nonumber\\
-\frac{1}{4}\beta|Z|^{-3}Z^3\dot{Z^{*}}^2
+\beta^3|Z|^{3}Z+\beta^2\lambda_1|Z|^{2}Z
+\lambda_2\beta|Z|Z+\lambda_3Z=0,
\label {1-tfeq01a} \\
\fl (II):\quad \tdot{Z}+(\lambda_1+4\beta Z)\ddot{Z}+3\beta
\dot{Z}^2
+(\lambda_2+3\beta \lambda_1Z+6\beta^2Z^2)\dot{Z}\nonumber\\
\qquad\qquad\qquad\qquad\qquad +(\lambda_1+\beta
Z)\beta^2Z^3+\lambda_2\beta Z^2+\lambda_3Z=0. \label {1-tfeq01b}
\end{eqnarray}
The associated real equation reads as
\begin{eqnarray}
\tdot{X}+(\lambda_1+4\beta X)\ddot{X}+3\beta \dot{X}^2
+(\lambda_2+3\beta \lambda_1X+6\beta^2X^2)\dot{X}\nonumber\\
\qquad\qquad\qquad\qquad\qquad +(\lambda_1+\beta
X)\beta^2X^3+\lambda_2\beta X^2+\lambda_3X=0. \label {1-tfeq01c}
\end{eqnarray}
Equation (\ref{1-tfeq01c}) is a generalized special case of the Chazy equation
XII \cite{Chand:07:nld} and the sub-cases of (\ref{1-tfeq01c}) has been studied
in detail in Refs. \cite{Chazy,Halburd,Cosgrove,Mugan,Euler,Euler:2005/06,
Chand5}.
The procedure to solve the above equation proceeds in a manner similar to that of the second order complex ODEs, and so we do not repeat them here. Similarly the procedure can be extended to the n-th order complex ODE case as well, generalizing the real ODE case discussed in ref.\cite{Chand:07:nld}

\section{Conclusion}
In this paper, we have developed a procedure to find integrable nonlinear complex ordinary differential equations by connecting the linear complex ODEs and nonlinear complex ODEs through two types of nonlocal transformations. We have also given a method to construct the general solution of the nonlinear complex ODEs from the linear complex ODEs. From the second type of nonlocal transformation, we have also pointed out that in certain cases the solution may have a multivalued nature.

\section{Acknowledgements}
The work of VKC and ML is supported by a Department of Science and Technology
(DST), Government of India, IRHPA research project. ML is also supported by a DAE
Raja Ramanna Fellowship and a DST – Ramanna program. The work of MS forms part of
a research project sponsored by the UGC and JHS is supported by a DST--FAST TRACK Young Scientist research project.

\section{Bibiliography}

\end{document}